# Quantitative use of electron energy-loss spectroscopy Mo-M$_{2,3}$ edges for the study of molybdenum oxides

L. Lajaunie*[1], F. Boucher, R. Dessapt and P. Moreau

*Institut des Matériaux Jean Rouxel (IMN) – Université de Nantes, CNRS, 2 rue de la Houssinière - BP 32229, 44322 Nantes Cedex 3, France*

*Corresponding author:* Tel: +34 976 762 783, E-mail address: luc.lajaunie@univ-nantes.fr

[1]*Present address:* Laboratorio de Microscopías Avanzadas (LMA), Instituto de Nanociencia de Aragón (INA), Universidad de Zaragoza, Campus Río Ebro C/ Mariano Esquillor s/n 50018 Zaragoza, Spain

**ABSTRACT**

Because of the large energy separation between O-K and Mo-L$_{2,3}$ edges, extracting precise and reliable chemical information from core-loss EELS analyze of molybdenum oxides has always been a challenge. In this regard Mo-M$_{2,3}$ edges represents an interesting alternative as they are situated close to the O-K edges. They should allow thus the extraction of a wealth of chemical information from the same spectra. However the main difficulty to overcome in order to work properly with these edges is the delayed maxima of the Mo-M$_{4,5}$ edges which hinders the automated background subtraction with the usual inverse power low function. In this study we propose another background subtraction method specifically designed to overcome this obstacle and we apply it to the study of MoO$_3$ and MoO$_2$. We are able to show that quantitative chemical information can be precisely and accurately determined from the joined analyze of O-K and Mo-$M_{2,3}$ edges. In particular k-factors are derived as a function of the integration window width and standard errors close to 2% are reported. The possibility to discriminate the two oxides thanks to chemical shifts and energy-loss near-edge structures is also investigated and discussed. Furthermore the $M_3/M_2$ ratios are derived and are found to be strongly dependent on the local chemical environment. This result is confirmed by multiplet calculations for which the crystal field parameters have been determined by *ab initio* calculations. The whole methodology as well as the conclusions presented in this paper should be easily transposable to any transitions metal oxides of the *4d* family. This work should open a new and easier way regarding the quantitative EELS analyses of these compounds.

**Highlights**

- EELS study of $MoO_3$ and $MoO_2$ compounds by using O-K and $M_{2,3}$ edges

- New method to subtract the background before *4d* transition metal $M_{2,3}$ edges

- Discussion of chemical shifts and ELNES of O-K and Mo-$M_{2,3}$ edges

- Mo-$M_{2,3}$ edges can be used for quantification with a great precision

- Good agreement between experimental and calculated $M_3/M_2$ ratios

# 1. INTRODUCTION

In the last few years, a lot of attention and efforts were dedicated to the nanoengineering of molybdenum oxides with control of size and shape. This has triggered the efficient and successful synthesis of an extraordinary variety of molybdenum oxide nanostructures with tunable morphologies including nanoplates, nanostars, nanowires, nanobelts, nanoflakes, comblike nanostructures, mesoporous nanowalls, nanorods and nanotubes [1–9]. Pioneering studies have shown that these nanostructures present enhanced physical and chemical properties when compared to their bulk counterparts and that they are promising candidates for a wide variety of technological and marketable applications including electronic and optoelectronic devices, gas-sensors, field-emitters, electrochemical supercapacitors and positive materials for Li-batteries [1,4,8,10–12]. This should open the way for an emerging field of research aiming at the characterization, comprehension and optimization of the properties of molybdenum-based nanostructures and associated nanodevices. The next challenge lies thus in the development of appropriate experimental tools for their characterization considering that as a strong need in this direction is expected. One key parameter influencing the chemical and physical properties of these compounds is their oxygen content that should be determined accurately and with enough spatial resolution [10,13,14].

Electron Energy-loss spectroscopy (EELS) performed in a transmission electron microscope (TEM) is a method of choice to perform quantitative chemical investigations at the nanoscale. Theoretical determination of the scattering cross-sections allow for the quantification of relative chemical concentrations of two elements A and B via Eq. (1) [15]:

$$\frac{N_A}{N_B} = \frac{I_A(\alpha,\beta,\Delta E)}{I_B(\alpha,\beta,\Delta E)} \bullet \frac{\sigma_B(\alpha,\beta,\Delta E)}{\sigma_A(\alpha,\beta,\Delta E)} \quad (1)$$

Where $N_i$ is the number of atoms per unit area, $I_i$ the integrated core-loss intensity within an energy windows of width $\Delta E$ starting close to the energy threshold of the energy edge and $\sigma_i(\alpha,\beta,\Delta E)$ the partial ionization cross-section integrated over a collection semi-angle β and corrected from the incident beam convergence semi-angle α. If the theoretical cross-sections are not known or not reliable, which is the case for Mo-$M_{2,3}$ edges, the k-factor can be experimentally determined from a standard of known A-B composition via Eq. (2) [16]:

$$k_{AB} = \frac{\sigma_B(\alpha,\beta,\Delta E)}{\sigma_A(\alpha,\beta,\Delta E)} = \frac{I_B(\alpha,\beta,\Delta E)}{I_A(\alpha,\beta,\Delta E)} \bullet \frac{N_A}{N_B} \quad (2)$$

It is important to note that these k-factors are dependent on many intrinsic and extrinsic parameters (such as α, β, $\Delta E$, and the methods used for background subtraction and multiple scattering deconvolution) and should be considered as a microscope and user dependent quantity. Generally speaking, EELS elemental quantification is a time consuming and user intensive procedure, which cannot be excessively automated. Beyond elemental quantification, many other quantitative and qualitative information can be deduced from a core-loss spectrum by studying, for instance, energy-loss near-edge structures (ELNES), integrated intensity ratios from $L_{2,3}$ edges as well as chemical shifts. The detailed analysis of all these aspects is far beyond the scope of this paper and the interested reader may refer to a numerous literature on this subject [15,17–19].

If EELS quantification is a well-established technique, quantitative EELS analysis of *4d* transition metals (TM) oxides remains a challenge so far. The $L_{2,3}$ edges of these TM oxides are situated at high-energies and cannot be used with confidence because of the excessively long dwell times and therefore unavoidable irradiation beam damages. For instance, the Mo-$L_{2,3}$ edges are located around 2500 eV and are too far away from the O-K edge (530 eV) to allow pertinent Mo valence determination and precise Mo/O elemental quantification from the same spectrum.

Compared to *3d*-TM, the EELS literature on *4d*-TM is thus scarce and focus mainly on the study of $M_{4,5}$ edges (involving transitions of *3d* electrons) which lies roughly between 150 and 300 eV. In 1987, Hofer has shown that $M_{4,5}$ edges can be used for elemental quantification of $MoO_3$ by determining the k-factors [20], and the same method was applied more recently to quantify niobium oxides [21,22]. The main drawback of using *4d*-TM-$M_{4,5}$ edges lies however in the delayed maximum that hinders the precise determination of the edge offset and thus has a detrimental effect on the precision of the elemental quantification. The presence of the C-K edge around 290 eV can also be an additional challenge in case of carbon contamination during the EELS experiment or study of carbon-contained materials.

**Fig. 1:** (Color online) **(a)** EELS spectra of a molybdenum oxide recorded with a dispersion of 0.30 eV/pixel to highlight the peaks position and shape of the Mo-$M_{4,5}$, Mo-$M_{2,3}$ and O-K edges **(b)** EELS spectra of a molybdenum oxide recorded with a dispersion of 0.20 eV/pixel highlighting the improper use of exponential and inverse power law functions to model the background continuum before the Mo-$M_{2,3}$ edge onset.

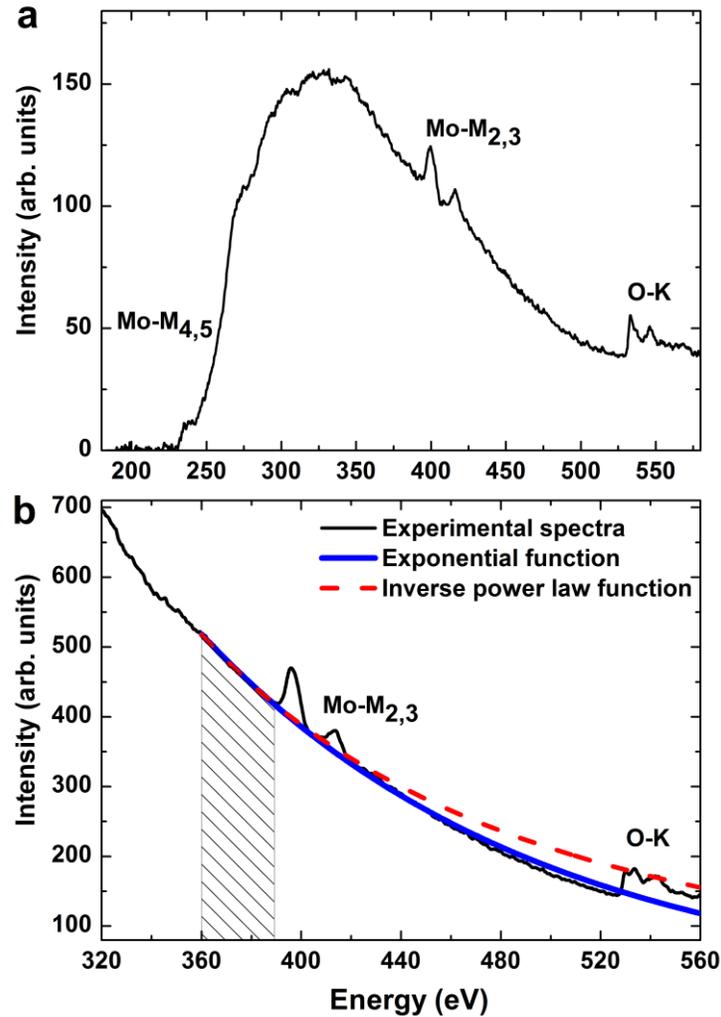

In order to overcome these difficulties, the *4d*-TM-$M_{2,3}$ edges (transitions involving electrons in *3p* orbitals) represent an interesting alternative. They are situated ideally close to the O-K edge (around 430 eV for Mo-$M_{2,3}$ edges for instance) and display sharp peaks similar to the $L_{2,3}$ white-line edges of TM and rare-earth compounds. To date, these edges are widely ignored by the EELS community and this is likely due to the delayed maxima of the $M_{4,5}$ edges (around 300 eV for Mo-$M_{4,5}$ edges) that appears as a broad rounded shoulder (Fig. 1a) and which hinders the automated background subtraction with the usual inverse power low function (Fig. 1b). Other approaches based on interpolation method [23] were also tested and found inappropriate to properly model the background before Mo-$M_{2,3}$ edges onset. However, the proximity between the O-K and *4d* $M_{2,3}$ edges is of great interest as it allows to collect them together while keeping a low experimental dispersion and thus avoiding a detrimental broadening effect on the energy resolution. A complete EELS study (going from ELNES analysis to elemental quantification and even determination of $M_3/M_2$ ratio) might thus be performed on the same spectrum in order to obtain a wealth of chemical information. To our knowledge, only one EELS study using O-K and Mo-$M_{2,3}$ edges has been reported so far in which the k-factor was determined from $MoO_3$ and was used for the elemental quantification of an unknown sample [24]. However, the authors did not give any details about the method they used for the background subtraction and, more importantly, the precision and the accuracy of their results was not discussed. Up to now, a quantitative and systematic EELS study of molybdenum oxides by means of $M_{2,3}$ edges is still missing. Due to the proximity of molybdenum with niobium in the periodic table, insights can also be gained from the EELS literature on niobium oxides [21,22,25]. For instance, Bach *et al*. modeled the background before the Nb-$M_{2,3}$ edges by using "predominantly an exponential function" and, from this, were able to derive the k-factors; however with a greater imprecision than the one determined from $M_{4,5}$ edges [21]. It is worth noting that, in the case of molybdenum oxides, an exponential function fails to model properly the background before the $M_{2,3}$ edges onset (Fig. 1b). Other studies, in which the background was simply not removed, have mainly focused on chemical shifts and O-K edge ELNES analyzes [22,25]. This might explain some inconsistency regarding a possible link between niobium valence states and $M_3/M_2$ intensity ratios.

In this paper, we explore the quantitative chemical information that can be derived from MoO$_3$ and MoO$_2$ compounds thanks to the joined analyze of Mo-M$_{2,3}$ and O-K edges. In section 2, the whole methodology is discussed. In particular, we present a new method to subtract the background, which is specifically designed to overcome the difficulties when working with *4d* transition metal M$_{2,3}$ edges. The details of the calculations based on density functional theory and multiplet approaches are also given. In section 3, the results regarding the ELNES, chemical shifts, elemental quantification and sensitivity of the M$_3$/M$_2$ ratios are presented and discussed. The method to subtract the background is validated by the great precision and accuracy of the results. Finally, this work gives a clear answer to the nature and the precision of the quantitative information that can be extracted by using these edges and when applying our new method for background subtraction.

## 2. MATERIALS AND METHODS
### 2.1 Materials and EELS acquisition parameters

α-MoO$_3$, the thermodynamically stable phase of the molybdenum trioxide in ambient conditions, and MoO$_2$ have been used in this study because they are well known simple molybdenum oxides which contain *4d*-metal ions with well-defined oxidation states (+IV and +VI, for MoO$_2$ and MoO$_3$ respectively), and in distinct chemical environments. Molybdenum trioxide α-MoO$_3$ is described in an orthorhombic unit cell (space group *Pbnm*) with the cell parameters *a* = 3.9624(1) Å, *b* = 13.860(2) Å, *c* = 3.6971(4) Å [26].The layered material is composed of [MoO$_3$] sheets stacked along the [010] direction, and which are held together by van der Waals interactions (Fig. 2a). In the structure, the Mo$^{VI}$ ion is linked to three crystallographically equivalent oxygen atoms in a strongly distorted octahedral environment (Mo-O bond lengths in the 1.67-2.33 Å range). The [MoO$_6$] octahedra are then condensed via edge and corner-sharing to give rise to the [MoO$_3$] sheets. Molybdenum dioxide MoO$_2$ has a rutile-like structure which is described in a monoclinic unit cell (space group $P2_1/c$) with the cell parameters $a$ = 5.6109(8) Å, $b$ = 4.8562(6) Å, $c$ = 5.6285(7) Å, $\beta$ = 120.95(1)° [27]. The Mo$^{IV}$ ion is linked to two crystallographically equivalent oxygen atoms in an octahedral environment. The [MoO$_6$] octahedra show very slight distortion (Mo-O bond lengths in the 1.97-2.07 Å range), and they are linked together by edge-sharing into infinite [MoO$_4$] chains running along the [100] direction (Fig. 2b). The chains are then condensed via corner-sharing into a three-dimensional MoO$_2$ network.

**Fig. 2:** (Color online) **(a)** Polyhedral representation of layered α-MoO$_3$: The [MoO$_3$] sheets are built upon distorted [MoO$_6$] octahedra linked together by edge- and corner-sharing (left). The layers are stacked along the $b$ axis (right). **(b)** Polyhedral representation of the rutile-like structure of MoO$_2$ (gold sphere: oxygen, blue octahedra: [MoO$_6$]).

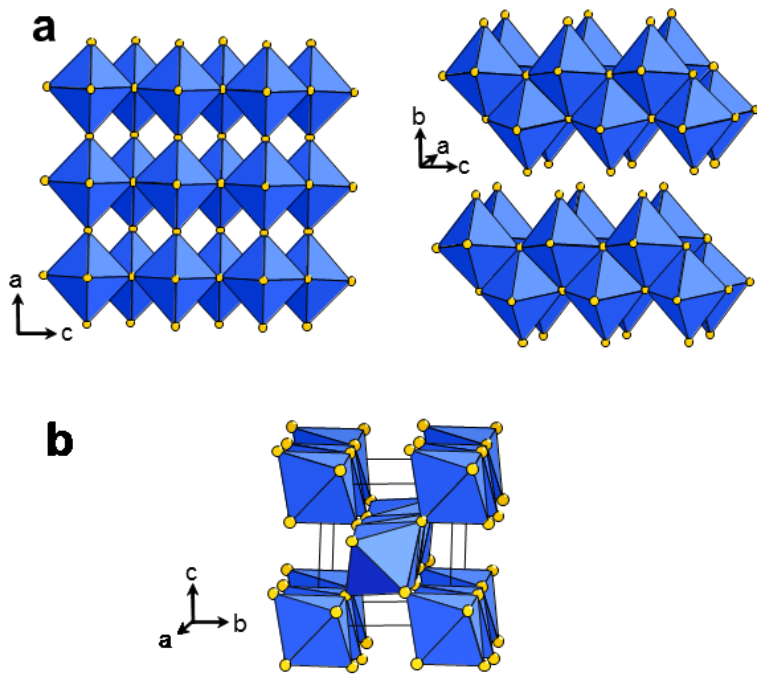

EELS spectra were acquired on commercial α-MoO$_3$ and MoO$_2$ powders using a Hitachi HF2000 TEM (100 kV) equipped with a cold field emission gun and a modified Gatan PEELS 666 spectrometer. Experiments were performed at liquid nitrogen temperature to minimize carbon contamination and irradiation beam damage which are known to be important in MoO$_3$ EELS experiments [24,28,29]. The acquisition time and the probe size for each core-loss were set to 20 s and 30 nm, respectively. The electron dose was measured in the same conditions with a Faraday cup and is equal to $2.7 \times 10^6$ e/nm$^2$. The energy resolution, measured as the full width at half maximum (FWHM) of the zero loss peak (ZLP), was 1.5 eV with an energy dispersion of 0.20 eV/pixel. This dispersion allowed the recording of the Mo-M$_{2,3}$ and O-K edges on the same spectra. Convergence and collection angle were 1.4 and 4.6 mrad, respectively. Thanks to this experimental setup ($\alpha/\theta_E = 0.7$, $\beta/\theta_E = 2.2$ for an energy-loss of 400 eV with an acceleration voltage of 100 kV), EELS spectra were acquired at the magic angle condition for the Mo-M$_{2,3}$ edges ($\beta_M/\theta_E=2.25$ for small $\alpha_M/\theta_E$ at 100 kV [30–33]) to avoid anisotropy effects playing a role in the determination of M$_3$/M$_2$ ratios. Each crystal was set slightly off zone axis to avoid channeling effects. At least 10 spectra per sample were acquired and processed to ensure the reproducibility of the results.

## 2.2 EELS data treatment

**Fig. 3:** (Color online) **(a)** method proposed in this paper to model the background continuum before the Mo-$M_{2,3}$ edge onset **(b)** the extracted spectrum after multiple scattering removal (thick black) and the tail of the cross-section of Mo-$L_{2,3}$ edges computed around 400 eV (thin green).

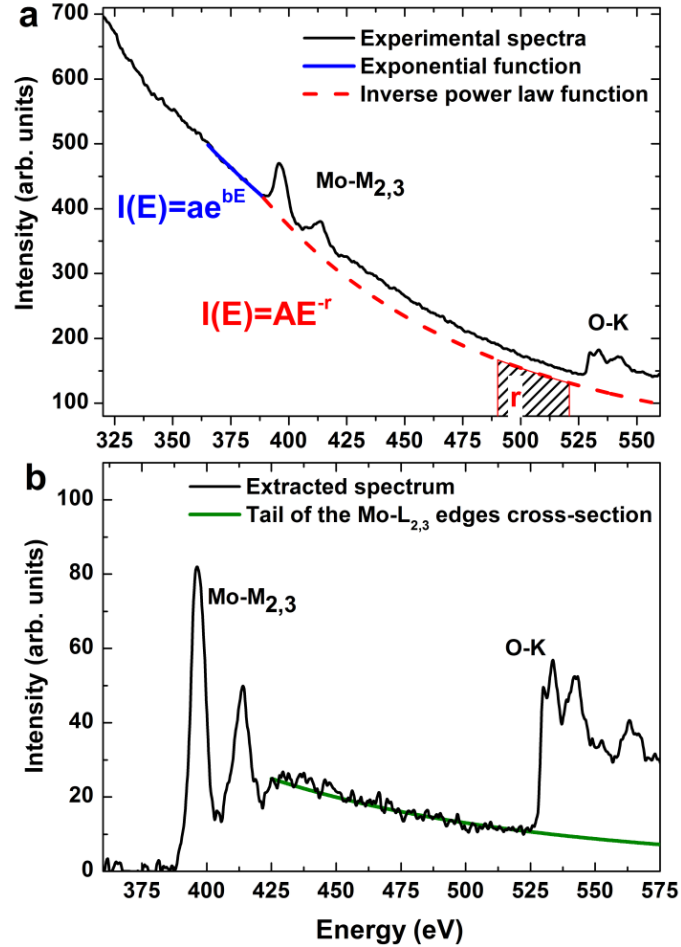

All spectra were first gain and dark count corrected and then deconvoluted by the ZLP using the PEELS program [34]. Background subtraction for the O-K edge was performed by modeling an inverse power law function $I = A_1 E^{-r_1}$ in the pre-edge energy window 490-520 eV, where E represents the energy loss and where both $A_1$ and $r_1$ are constants. The extrapolated background was then subtracted as illustrated in Fig. 3a. Starting from the $M_3$ edge onset (~389 eV) the background was modeled by using an inverse power law function $I = A_2 E^{-r_1}$ where the $r_1$ parameter is the same than the one determined for the O-K edge. The main objective of this method is to avoid the detrimental effect of the Mo-$M_{4,5}$ edges on the background subtraction. In addition, before the Mo-$M_3$ edge onset, the background was modeled by using an exponential function ($ae^{bE}$, where $a$ and $b$ are constants) for the sole purpose to have a pre-edge qualitative indicator of the signal/noise ratio. In order to reach convergence in an efficient manner, these two fits were achieved by using specific routines based on Nelder-Mead algorithms (unconstrained nonlinear optimization) [35].

The multiple scattering was then removed by Fourier-ratio deconvolution with the low-loss spectrum obtained for exactly the same region of the sample [15]. The resulting spectrum is displayed in Fig. 3b. In addition, the theoretical cross-section of Mo-$L_{2,3}$ edges was calculated by using the parameterization of the generalized oscillator strength [36] thanks to the PEELS program [34]. For this purpose, the cross-section was calculated by constraining the maxima of the Mo-$L_{2,3}$ edges to be close to the experimental maxima of the Mo-$M_{2,3}$ edges. The tail of this calculated cross-section is displayed in Fig. 3b. As it can clearly be seen from this figure, there is an excellent agreement between the tails of the cross-section and the tail of the $M_{2,3}$ edges after background and multiple scattering removal. This demonstrates that the method used to remove the background allows to recover an absorption edge which decreases in a coherent manner with what would be expected in similar conditions of incident beam energy, energy position and collection and convergence angles.

To determine the feasibility of elemental quantification when using Mo-$M_{2,3}$ edges, the k-factors $k_{MoO}$ were determined from $MoO_3$ and $MoO_2$ samples according to Eq. (2). The width of the energy windows ($\Delta E$) was varied from 5 to 60 eV (by step of 5 eV) to check on the influence of this parameter on the quantification results. Due to the low energy dispersion, larger energy window than 60 eV could not be achieved. To determine the $M_3/M_2$ intensity ratios, the maximum of the $M_3$ edges were first aligned to the same energy-loss (398 eV) to minimize the systematic error from peak positions and shapes. Following closely the work of Daulton and Little [37], a two-steps "zero-slope" function with the step onsets occurring at the white-line maximums was build and was further convoluted by a Gaussian function with a FWHM of 1.5 eV to match the experimental resolution. According to the authors, the "zero-slope" method yields less scattered results than the method used for instance by Pearson *et al*. [38], in which the slope follows the decrease of the cross-section of the considered edges. The ratio of the steps heights was set as 2:1 to reflect the multiplicity of the $3p_{1/2}$ and $3p_{3/2}$ initial states and the intensity function was adjusted to match the minimum intensity of the experimental spectrum between 420 and 425 eV. The two-steps function was then subtracted from the experimental spectra and the $M_3/M_2$ intensity ratios were determined by area integration from 399 to 407 eV and 407 to 423 eV for the $M_3$ and $M_2$ edges, respectively (Fig. 4). A great care was taken to check that no correlation exists between the experimental $M_3/M_2$ intensity ratios and the corresponding relative thicknesses (deduced from low-loss spectra) to rule out an improper multiple scattering deconvolution and/or the influence of irradiation beam damages [39].

**Fig. 4:** (Color online) Method used for the determination of the $M_3/M_2$ ratios. A two-steps function was subtracted and the ratios were determined by integration of the areas under the peaks.

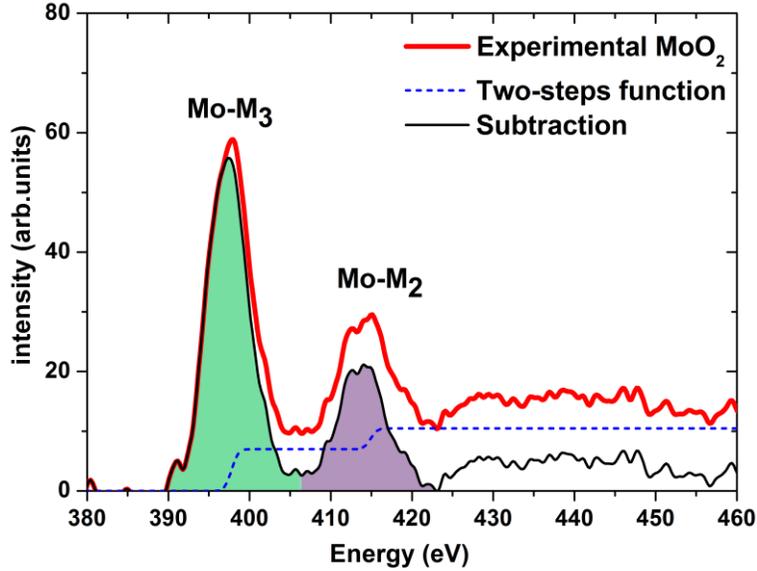

**2.3 Multiplet and *ab initio* calculations**

In order to check on the accuracy of the method proposed in this paper, experimental intensity ratios were also compared to theoretical one derived from multiplet calculations by using the CTM4XAS program [40]. Calculations were performed by reducing Slater integrals to 80% of their calculated atomic values. A 0.5 eV Lorentzian broadening was added to take into account life-time broadening and a 1.5 eV Gaussian broadening was also added corresponding to the experimental energy resolution. The crystal field parameters (*10 Dq*), which were found to strongly affect calculated intensity $M_3/M_2$ ratios, were obtained from *ab initio* calculations based on the Density Functional Theory [41]. For this purpose, the ground-state electronic structures of α-$MoO_3$ an $MoO_2$ were determined with the all electron code WIEN2k [42] in the Perdew-Burke-Ernzerhof parameterization of the generalized gradient approximation (PBE-GGA). The structural data of α-$MoO_3$ and $MoO_2$ were taken from the work of Lajaunie *et al.* and Bolzan *et al.*, respectively [43,44]. Muffin-tin radii ($R_{MT}$) were set to 1.96 Bohr radius for Mo and 1.7 Bohr radius for O. The self-consistencies on electronic density were obtained with a plane wave energy cutoff defined by $R_{MT} \times K_{MAX} = 7$ and by using (8×2×8) and (8×8×8) $k$-point meshes for $MoO_3$ and $MoO_2$, respectively. The *10 Dq* were then determined from the band structures calculations by taking the average separation between the $t_{2g}$ and $e_g$ bands and were found to be equal to 3.5 eV for $MoO_3$ and 4.5 eV for $MoO_2$. Finally, theoretical $M_3/M_2$ ratios were derived from the calculated multiplet spectra by area integration in a similar manner than the one used for the experimental ratios.

## 3. RESULTS AND DISCUSSION

### 3.1 ELNES and chemical shifts

Electron-energy-loss near edge structures (ELNES) originate from the electron transitions from core orbitals to unoccupied bands. These spectral features can be envisaged in a first-order approximation as an image of a momentum and atomic resolved projection of the unoccupied density of states. They are thus sensitive to the chemical bonding. Fig. 5a shows the ELNES of the O-K edge for $MoO_3$ and $MoO_2$. Both spectra are similar to those previously reported by EELS and XAS studies on the same compounds [24,45–50]. This shows that the dwell time of the EELS acquisition was set short enough to avoid the reduction of the samples under the electron beam as it was previously reported on molybdenum oxides [24].

**Fig. 5:** (Color online) EELS spectra recorded on MoO$_3$ and MoO$_2$ samples showing the ELNES of the O-K edges **(a)** and Mo-M$_{2,3}$ edges **(b)**.

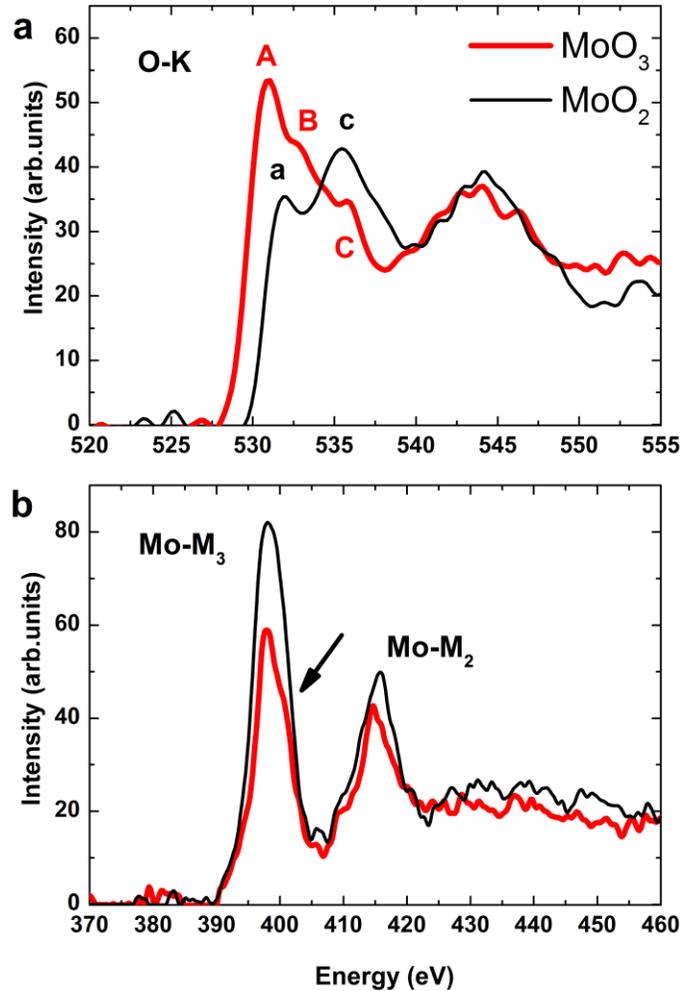

Fig. 5a clearly shows that MoO$_3$ and MoO$_2$ can be distinguished thanks to the spectral shape of their O-K edges over an energy range of 10 eV above the absorption threshold. In particular, the O-K edge of MoO$_3$ presents a first strong peak (labeled *A* Fig. 5a and with an absolute position of 531.2 ± 0.3 eV) followed by two small shoulders (labeled *B* and C and situated around 532.5 eV and 534.5 eV, respectively) whereas MoO$_2$ presents two peaks (labeled *a* and *c* in Fig. 5a and situated at 531.1 eV ± 0.5 and 534.5 ± 0.4 eV, respectively) among which the second peak is the most intense. Going beyond this qualitative description would require a more complete experimental and theoretical study to link the O-K edge fine structures of these two phases with their crystallographic structures as it have been done for other compounds [51]. Such study is still lacking even if previous studies on $MoO_3$ and $MoO_2$ interpreted the peaks near the absorption threshold as resulting from transitions of oxygen *s* core states to oxygen *p* states hybridized with the transitions metal $t_{2g}$ and $e_g$ states [41,50,52]. The ELNES of Mo-$M_{2,3}$ edges for the two compounds are shown in Fig. 5b. Except from a double-peak structure for the $M_3$ edge of $MoO_3$ (highlighted by an arrow in Fig. 5b) no clear spectral feature is distinguishable on the two spectra. The smallest features around the $M_2$ edges are not reproducible and result from random noise rather than fine-structures. It is worth noting that except a double peak-structure in the $M_2$ edge of $MoO_3$, XAS studies with considerably better energy resolutions (around 0.1-0.2 eV) do not reveal more specific features [50,53,54]. Absolute chemical shifts and energy differences can also be used as a fingerprinting method to discriminate compounds with different oxidation states [15,19,55–57] although the measurement of absolute positions is quite tricky due to the TEM instabilities and often requires specific experimental setup or software during the data acquisition [19,58,59]. A better precision can generally be obtained by analyzing the difference in energy between two ELNES features. The question is thus to determine which ELNES quantity behaves most sensitively with a change of valence state.

**Tab. 1:** Energy position of the maximum of the Mo-$M_3$ edge, energy difference between the maxima of the Mo-$M_3$ and Mo-$M_2$ edges and energy difference between the maximum of the Mo-$M_3$ edge and the inflexion point of the O-K edge.

|  | **Mo-$M_3$ (eV)** | **$M_3$<->$M_2$ (eV)** | **Mo-$M_3$<->O-K (eV)** |
|---|---|---|---|
| **MoO$_3$** | 397.8 ± 0.4 | 16.8 ± 0.3 | 131.6 ± 0.2 |
| **MoO$_2$** | 397.7 ± 0.5 | 16.5 ± 0.4 | 132.2 ± 0.2 |

Tab. 1 presents the absolute position of the $M_3$ maximum for the two oxides, the difference in energy between the two M edges maxima and finally, the difference in energy between the $M_3$ maximum and the inflexion point of the O-K edge (determined by the change of sign of the second derivative). It is clear that MoO$_3$ and MoO$_2$ cannot be distinguished from the first ELNES quantity, the average absolute position of the $M_3$ maximum being identical. The same statement can be made for the second ELNES quantity since the difference falls in the error range. This is mostly due to the low signal/noise ratio around the M edges which hinders the precise determination of the $M_2$ maximum. The measurements of the energy difference between the $M_3$ maximum and the inflexion point of the O-K edge reveal a clear increase of 0.6 eV going from MoO$_3$ to MoO$_2$. With respect to the standard errors, this last result is statistically significant. This evolution is consistent with the metallic and semiconductor nature of MoO$_2$ and MoO$_3$, respectively. More quantitative statements would however require comparison with calculations based on many-body approaches which are beyond the scope of this paper. In addition, the sensitivity of this ELNES quantity with this valence state is rather weak and other possibilities such as elemental quantification and $M_3$/$M_2$ ratios are thus worth exploring.

## 3.2 Elemental quantification and $M_3/M_2$ ratios

**Fig. 6:** (Color online) **(a)** k-factors for $MoO_3$ (red squares) and $MoO_2$ (black circles) determined from the Mo-$M_{2,3}$ and O-K edges as a function of the energy window, $\Delta E$. **(b)** Standard errors (95% confidence interval) on the determination of the k-factors for $MoO_3$ (red squares) and $MoO_2$ (black circles) and relative difference to the mean defined by $\dfrac{k_{MoO} - \overline{k_{MoO}}}{k_{MoO}}$ (blue diamonds).

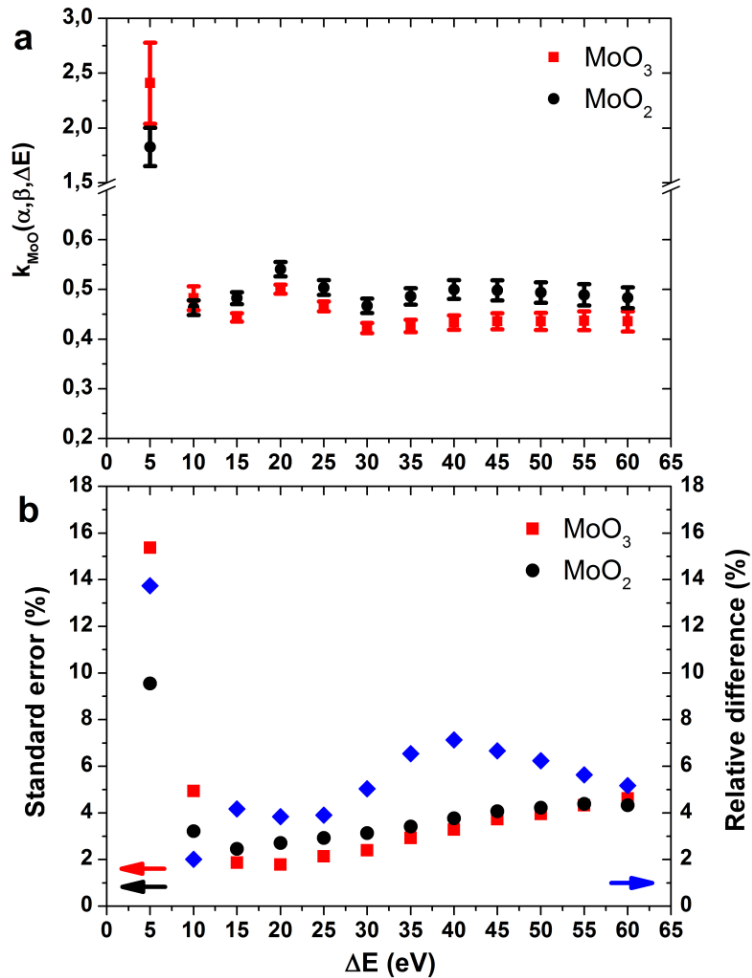

One of the main purpose of this paper is to determine if a reproducible and accurate elemental quantification can be achieved by using Mo-$M_{2,3}$ and O-K edges. It is generally assumed that large energy windows should be avoided because they tend to maximize errors due to background extrapolation and thus compromise the precision of the quantification. On the other hand, it is also advised to avoid small energy windows because of the differences of near-edge fine structures which compromise the accuracy of the quantification. To clarify this point, we have determined the k-factors and the corresponding standard deviation (95% confidence interval) as a function of the energy width in Fig. 6. Except for the unreasonable energy window of 5 eV, all the values of $k_{MoO}$ are scattered close to 0.45 (Fig. 6a), thus showing that the partial cross-section obtained from the Mo-$M_{2,3}$ edge is roughly twice as large as that obtained from the O-K edge. The standard errors are the highest for a width of 5 eV (around 15% and 10% for $MoO_3$ and $MoO_2$, respectively), reach a minimum close to 2% for energy windows of 15 and 20 eV for $MoO_2$ and $MoO_3$ respectively and, from there, increase slowly with the window width to reach a value around 4.5% for a width of 60 eV (Fig. 6b). Once averaged, the results obtained from the two compounds show a local minimum of the standard error for an energy window of 15 eV ($k_{MoO}$ = 0.46 ± 0.01 i.e., 2.2% of standard errors). Compared to literature, this result is more precise than that obtained from Bach *et al.* by using Nb-$M_{2,3}$ edges on niobium oxides (±7%) and presents a similar precision than their k-factors determined from Nb-$M_{4,5}$ edges (±2.1%) [21]. Olszta *et al.* determined the k-factors from the Nb-$M_{4,5}$ edges solely and reached a precision slightly larger than 4% [22]. The precision and the accuracy of these results validate the method we used to subtract the background. In order to determine the accuracy of our results (and to check if the average values of the two k-factors can be used for quantification), the relative difference to the mean is also plotted in Fig. 6.(b). This quantity is defined as $\dfrac{k_{MoO} - \overline{k_{MoO}}}{k_{MoO}}$, where $k_{MoO}$ is the k-factor determined either from the $MoO_2$ or the $MoO_3$ sample and where $\overline{k_{MoO}}$ is the mean value between the k-factors determined from the two samples. The relative difference to the mean presents strong variations depending on the energy window. The highest difference (14%) occurs with a width of 5 eV whereas the best accuracy (2%) on the determination of the k-factors is found for a width of 10 eV. A plateau with values around 4% is also found for energy width of 15, 20 and 25 eV. To summarize and illustrate the effects of accuracy and precision of the quantification, Fig. 7 shows the results of quantification for the two samples by using the average k-factors. The standard errors on quantification are evaluated by taking into account the errors on determination of the k-factors and integrated intensities. The highest accuracy is achieved with a 10 eV window width ($N_{Mo}/N_O$ = 0.33 ± 0.03 (± 9%) and 0.51 ± 0.04 (± 8%) for $MoO_3$ and $MoO_2$, respectively) whereas the highest precision is achieved with a 15 eV window width ($N_{Mo}/N_O$ =0.35 ± 0.01 (± 3%) and 0.48 ± 0.02 (± 4%) for $MoO_3$ and $MoO_2$, respectively). These two windows are thus the best choices for elemental quantification and should be selected accordingly with the desired objective.

**Fig. 7:** (Color online) Results of the quantification for the $MoO_3$ and $MoO_2$ samples by using the average k-factors. The expected stoichiometries are given by the blue lines.

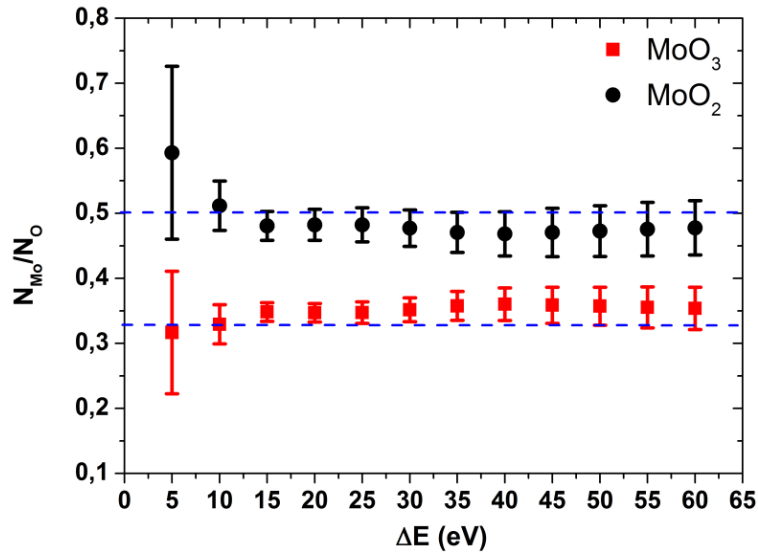

Another interrogation remains concerning the sensitivity of the *4d*-TM -$M_{2,3}$ edges with the chemical environment of the metal. On this subject, contradictory results can be found in the EELS literature related to niobium oxides [22,25]. In particular, by using a variant of the Pearson's method [38], Olszta et *al.* reported a correlation between the normalized white-line intensities and the *4d* occupancy. However the authors did not discuss the accuracy of their method (in which the background subtraction before the Nb-$M_{2,3}$ edges onset was not performed), and when applying the same method to $MoO_3$ $M_{2,3}$ edges, we find a statistical error of nearly 30% on the sum of the white-line intensities. Due to the strong similarities between *4d*-$M_{2,3}$ and *3d*-$L_{2,3}$ edges related to their white-line nature (transitions from spin orbit split initial states to quasi-localized final states), valuable insights can be gained again from the EELS and XAS literature related to *3d*-$L_{2,3}$ edges. The most popular and common EELS method to link the $L_{2,3}$ edges with the chemical environment of a *3d*-TM is without any doubt the $L_3/L_2$ integrated intensity ratio. This method is issued from the interrogations which were raised in the 80's about the deviation between the experimental branching ratio and the statistical one (*i.e.* as expected from the degeneracy of the *2p* states into $2p_{1/2}$ and $2p_{3/2}$ levels) [60–62]. Nowadays the origins of this deviation are still under investigation [63,64]. However it can be stated that the $L_3/L_2$ ratio is not only sensitive to the valence state of the metal [37,55] but rather to a complex mix between the local electronic structure of the cation, its local environment and, to some extent, bandstructure effects [65,66]. In this regard, this method is extremely interesting to track down subtle chemical changes on the local coordination but is extremely difficult to interpret without multiplet and/or *ab initio* calculations and is thus generally used as a fingerprinting method. Many variants of this method have been proposed and applied successfully on a large variety compounds [18,19,29,37,55].

**Fig. 8:** (Color online) Crystal field multiplet calculation spectra for Mo$^{6+}$ (thick red line) and Mo$^{4+}$ (thin black line) calculated in octahedral environment.

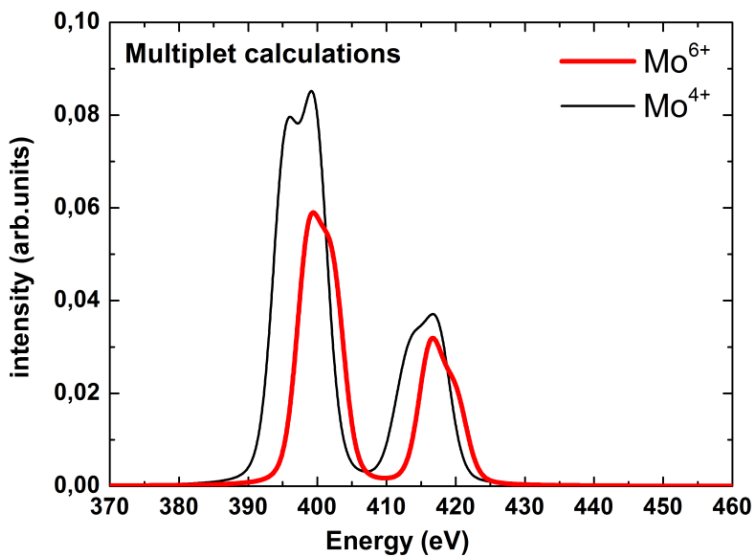

For MoO$_3$ and MoO$_2$, the experimental M$_3$/M$_2$ ratios were found to be equal to 1.91 ± 0.05 (±3%) and 2.20 ± 0.04 (±2%), respectively. Thanks to the method used for the background extraction, a high precision is found on the determination of the ratios. This result is of importance since it shows that the M$_3$/M$_2$ ratios are sensitive to the chemical environment of the *4d* cation and that they can be used to make a clear distinction between the two oxides. In order to check on the validity of these numbers, the M$_3$/M$_2$ ratios were also determined from multiplet calculated spectra (Fig. 8) for which the crystal field parameters have been previously estimated by DFT calculations. From a qualitative point of view, the calculated intensities match the intensities of the experimental spectra (Fig. 5b). For MoO$_3$ and MoO$_2$, the calculated M$_3$/M$_2$ ratios are equal to 1.99 and 2.26, respectively and are thus really close to the experimental ratios. This results assets our experimental methodology and it also implies that the "real" background under the M edges has been properly estimated.

## 4. CONCLUSION

We have proposed another background subtraction method specifically designed to circumvent the difficulties arising when working with the $M_{2,3}$ edges of the *4d* transition metals oxides. This method was applied to $MoO_3$ and $MoO_2$ and allowed us to explore new possibilities given by the simultaneous analysis of O-K and Mo-$M_{2,3}$ edges in term of ELNES differences, chemical shifts, elemental quantification and $M_3/M_2$ ratio. Firstly, the two molybdenum oxides can be distinguished qualitatively thanks to the spectral shape of their O-K edges over an energy range of 10 eV. The difference in energy between the maximum of the $M_3$ edge and the inflexion point can also be used since it was found to increase significantly going from $MoO_3$ to $MoO_2$. Secondly, the $M_{2,3}$ edges can be used for elemental quantification and a great care was taken to determine the accuracy and precision of the k-factors as a function of the energy window used for the area integration. In particular, the best precision of the determination of k-factors is realized with an energy window of 15 eV and is close to 2%. Furthermore when comparing the k-factors determined from $MoO_3$ and $MoO_2$, the same energy window yields an accuracy of 4%. This shows that Mo-$M_{2,3}$ edges can be used for elemental quantification with a precision equal or even better than those previously reported on *4d* transition metal oxides with $M_{4,5}$ edges. Furthermore the $M_3/M_2$ ratios can also be used to track down subtle chemical changes since they were found to be sensitive to the chemical environment of the *4d* cation. In addition, the validity of the numbers extracted from the $M_3/M_2$ analysis was confirmed by multiplet calculations. This work thus demonstrates the ability to obtain a wealth of precise and accurate chemical information on molybdenum oxides from the conjugated analyzes of O-K and Mo-$M_{2,3}$ edges when the background is properly extracted thanks to the method proposed in this paper. It should also open interesting opportunities for the EELS studies of a large variety of materials as it is directly transposable to the whole family of *4d* transition metal oxides.

## ACKOWLEDGMENTS

We would like to thank Pr. F.M. De Groot for his help and discussion in the use of the CTM4XAS program.
## REFERENCES

[1] T. Brezesinski, J. Wang, S.H. Tolbert, B. Dunn, Ordered mesoporous alpha-MoO3 with iso-oriented nanocrystalline walls for thin-film pseudocapacitors, Nat Mater. 9 (2010) 146–151. doi:10.1038/NMAT2612.

[2] D. Chen, M. Liu, L. Yin, T. Li, Z. Yang, X. Li, et al., Single-crystalline MoO3 nanoplates: topochemical synthesis and enhanced ethanol-sensing performance, J. Mater. Chem. 21 (2011) 9332–9342.

[3] S. Hu, X. Wang, Single-walled MoO3 nanotubes, J. Am. Chem. Soc. 130 (2008) 8126–8127.

[4] A. Khademi, R. Azimirad, A.A. Zavarian, A.Z. Moshfegh, Growth and field emission study of molybdenum oxide nanostars, J. Phys. Chem. C. 113 (2009) 19298–19304.

[5] L. Mai, F. Yang, Y. Zhao, X. Xu, L. Xu, B. Hu, et al., Molybdenum oxide nanowires: synthesis & properties, Mater Today. 14 (2011) 346–353.

[6] A. Okada, M. Yoshimura, K. Ueda, Fabrication of comblike nanostructures using self-assembled cluster arrays of molybdenum oxides, Appl. Phys. Lett. 90 (2007) 203102–203102.

[7] L.X. Song, J. Xia, Z. Dang, J. Yang, L.B. Wang, J. Chen, Formation, structure and physical properties of a series of alpha-MoO3 nanocrystals: from 3D to 1D and 2D, CrystEngComm. 14 (2012) 2675–2682. doi:10.1039/c2ce06567c.

[8] Z. Wang, S. Madhavi, X.W. (David) Lou, Ultralong alpha-MoO3 Nanobelts: Synthesis and Effect of Binder Choice on Their Lithium Storage Properties, J Phys Chem C. 116 (2012) 12508–12513. doi:10.1021/jp304216z.

[9] B. Yan, Z. Zheng, J. Zhang, H. Gong, Z. Shen, W. Huang, et al., Orientation Controllable Growth of MoO3 Nanoflakes: Micro-Raman, Field Emission, and Birefringence Properties, J. Phys. Chem. C. 113 (2009) 20259–20263.

[10] S. Balendhran, J. Deng, J.Z. Ou, S. Walia, J. Scott, J. Tang, et al., Enhanced Charge Carrier Mobility in Two-Dimensional High Dielectric Molybdenum Oxide, Adv. Mater. 25 (2013) 109–114.

[11] N. Illyaskutty, S. Sreedhar, H. Kohler, R. Philip, V. Rajan, V.M. Pillai, ZnO-Modified MoO3 Nano-Rods,-Wires,-Belts and-Tubes: Photophysical and Nonlinear Optical Properties, J. Phys. Chem. C. 117 (2013) 7818–7829.

[12] P. Meduri, E. Clark, J.H. Kim, E. Dayalan, G.U. Sumanasekera, M.K. Sunkara, MoO3–x Nanowire Arrays As Stable and High-Capacity Anodes for Lithium Ion Batteries, Nano Lett. 12 (2012) 1784–1788.

[13] I. Navas, R. Vinodkumar, V.P.M. Pillai, Self-assembly and photoluminescence of molybdenum oxide nanoparticles, Appl Phys Mater Sci Process. 103 (2011) 373–380. doi:10.1007/s00339-011-6345-9.
25